\documentclass[twoside,a4paper]{IEEEtran}

\usepackage[IEEEtran]{research18}

\begin{document}

\title{State-Dependent DMC with a Causal Helper}

\author{%
    Amos Lapidoth and Ligong Wang\thanks{The authors are with the Department of Information Technology and Electrical Engineering, ETH Zurich, 8092 Zurich, Switzerland (e-mail: \mbox{lapidoth@isi.ee.ethz.ch}; \mbox{ligwang@isi.ee.ethz.ch}).}}

\maketitle

\begin{abstract}
  A memoryless state sequence governing the behavior of a memoryless
  state-dependent channel is to be described causally
  to an encoder wishing to communicate over said channel. Given the
  maximal-allowed description rate, we seek the description that
  maximizes the Shannon capacity. It is shown that the maximum need not
  be achieved by a memoryless (symbol-by-symbol) description.
  Such descriptions are, however, optimal when the receiver is cognizant
  of the state sequence or when the description is allowed to depend on
  the message. For other cases, a block-Markov scheme with backward
  decoding is proposed.
\end{abstract}


\section{Introduction and Problem Setup}
The impact of state information on the capacity of a state-dependent
discrete memoryless channel (SD-DMC) is well understood. State
information at the receiver can be accounted for by appending it to
the output, and the impact of state information at the transmitter
depends on its timing: if it is provided strictly causally, it has no
impact on capacity; if causally, then the capacity is as given by
Shannon \cite{shannon58}; and if noncausally, then as given by
Gel'fand and Pinsker \cite{gelfandpinsker80_3}. Less studied is how the
state information should be conveyed to the encoder when rate
restrictions preclude its precise description. We address this issue
here by studying the design and impact of rate-limited causal state
descriptions to the encoder.

We account for the rate constraint by requiring that the time-$i$
assistance provided to the encoder take value in some fixed set
$\set{T}$, whose cardinality $\card{\set{T}}$ is typically smaller
than that of the state alphabet $\set{S}$. (When $\card{\set{T}} \geq
\card{\set{S}}$ we are back to Shannon's causal state information,
because the helper can then describe the state precisely.)  We assume
throughout that 
\begin{equation}\label{eq:assumptionT}
        \card{\set{T}} \geq 2
\end{equation}
because, otherwise, the description is of no help. We refer to $\log
\card{\set{T}}$ as the \emph{description rate}.

This way of accounting for rate restrictions is quite rigid: it allows
for neither variable-length state descriptions nor for time sharing
between fine and coarse quantizations. Precluding these techniques
sharpens some of our conclusions.

It should be emphasized that causality does not imply that the helper
must describe the state sequence ``symbol-by-symbol,'' i.e., that the time-$i$
assistance $T_{i} \in \set{T}$ be determined by the time-$i$ state
$S_{i}$: the time-$i$ assistance may depend on the entire state sequence up
to time~$i$, namely, $S^{i}$.

   The fact that we only consider memoryless channels with independent and 
identically ditributed (IID) states
 might lead one to suspect that symbol-by-symbol helpers are
 optimal. Lending credence to this suspicion might be that, when the
 causal state information is perfect (i.e., when $\card{\set{T}} \geq
 \card{\set{S}}$), Shannon strategies achieve capacity, and those
 ignore the past states and set the time-$i$ channel input
 $X_{i}(m,S^{i})$ to be a function of the message $m$ and the time-$i$
 state $S_{i}$ only. But this is not the case. As we shall see in
 Example~\ref{ex:from_hell} ahead, causal helpers can outperform the
 best symbol-by-symbol helpers. This example will motivate us to
 propose a block-Markov communication scheme that can outperform all
 symbol-by-symbol schemes.

 Symbol-by-symbol descriptions are, however, optimal in some special
 cases, e.g., when the state is known perfectly to the receiver
 (Theorem~\ref{thm:psi_at_decoder}). They are also optimal when the
 helper is cognizant of the message (Theorem~\ref{thm:hkm}).  They
 are 
 effective in the sense that if some positive rate is achievable with
 a general rate-limited causal helper, then a positive rate is also
 achievable with a symbol-by-symbol helper. In fact, subject to \eqref{eq:assumptionT},
 symbol-by-symbol helpers can achieve positive communication rates
 whenever the SD-DMC is of positive capacity when its state is
 revealed perfectly to both encoder and decoder
 (Theorem~\ref{thm:positive}).

We only present results for point-to-point channels. State information
is, of course, of great importance also in multi-user settings, but
those are more complicated and are not even fully understood when the
helper's rate suffices for perfect state description. In such settings
strictly-causal state information may be helpful, and Shannon
strategies are sub-optimal \cite{lapidothsteinber13common,lapidothsteinberg13double,lisimeoneyener13}. For much
more on state information in communication systems, see
\cite{keshetsteinberg08}. Related to our work is
\cite{rosenzweigsteinbergshamai05} which---subject to the assumption
of perfect receiver state information---analyzes a system comprising a
symbol-by-symbol and a noncausal helper to the transmitter.

\subsection{The channel model}

We consider a state-dependent discrete memoryless channel 
with finite input, output, state, and description alphabets $\set{X}$, $\set{Y}$,
$\set{S}$, and $\set{T}$, respectively.
The states $S_1,S_2,\ldots$ are IID $\sim P_{S}$, where $P_{S}$ is
some given probability mass function (PMF) on
$\set{S}$. The channel is memoryless, and we denote its law $W(y|x,s)$: given the channel
input $x \in \set{X}$ and the channel state $s \in \set{S}$, the probability of the channel output
being $y \in \set{Y}$ is $W(y|x,s)$ 
irrespectively of past inputs, states, or outputs. A fortiori, for every
$x \in \set{X}$ and $s \in \set{S}$, the transition law $W(y|x,s)$ is
nonnegative and $\sum_{y\in\set{Y}}W(y|x,s)=1$.

When communicating with blocklength $n$, the transmitted message $m$
is chosen from the set of messages $\set{M}=\{1,\ldots,2^{nR}\}$,
where $R$ is the communication rate.

When the encoder is provided with perfect causal state information---a
setting to which we refer as the ``Shannon setting,'' because
this is the setting studied in \cite{shannon58}---a blocklength-$n$
encoder consists of $n$ mappings
\begin{equation}
f_i\colon \set{M}\times\set{S}^i \to \set{X},\quad (m,s^i)\mapsto x_i,\quad i=1,2,\ldots, n
\end{equation}
with the understanding that the time-$i$ channel input $X_{i}$
that the encoder produces in order to convey the message $m$ when the
present and past states are $S^{i}$ is $X_{i} = X_{i}(m,S^{i}) =
f_{i}(m,S^{i})$.

When the encoder is provided with \emph{perfect strictly-causal} state
information, these mappings take the form
\begin{equation}
f_i\colon \set{M}\times\set{S}^{i-1} \to \set{X},\quad (m,s^{i-1})\mapsto x_i, \quad i=1,2,\ldots, n
\end{equation}
with $X_{i}$ now being a function of the message and past (and not present) states,
i.e., $X_{i} = X_{i}(m,S^{i-1}) = f_{i}(m,S^{i-1})$.

The focus of our work, however, is the \emph{causal helper
setting}, where the helper consists of a sequence of mappings
\begin{equation}\label{eq:h_general}
h_i\colon \set{S}^i \to \set{T},\quad s^i \mapsto t_i,  \quad
i=1,2,\ldots, n 
\end{equation}
with the understanding that the time-$i$ help\footnote{We use ``help,'' 
``assistance,'' and ``description'' interchangeably.} $T_{i}$ that is provided to
the encoder is 
$T_{i} = T_{i}(S^{i}) = h_{i}(S^{i})$. The time-$i$ channel input
produced by the encoder to convey the message $m$ is then a function
of $m$ and the present-and-past help $T^{i}$. The encoder thus
consists of a sequence of mappings
\begin{equation}\label{eq:encoder_helped_causally}
f_i\colon \set{M}\times \set{T}^i\to \set{X},\quad (m,t^i)\mapsto x_i,  \quad
i=1,2,\ldots, n 
\end{equation}
with the understanding that the time-$i$ channel input $X_{i}$
produced by the encoder to convey the message $m$, after having
received the present-and-past help $T^{i}$, is $X_{i} = X_{i}(m,T^{i})
= f_{i}(m,T^{i})$. A special kind of helper is the
\emph{symbol-by-symbol} helper whose time-$i$ help $T_{i}$ is a
function only of the time-$i$ state $S_{i}$, i.e., it is a helper
where the mappings \eqref{eq:h_general} have the form
\begin{equation}
\label{eq:sbs}
h_i\colon \set{S}\to\set{T},\quad s_i\mapsto t_i.
\end{equation}

In all the above cases, a decoder is a mapping
\begin{equation}
g\colon \set{Y}^n\to\set{M},\quad y^n\mapsto \hat{m}
\end{equation}
with the understanding that, upon receiving the channel output
sequence $Y^{n}$, the decoder produces the message $g(Y^{n})$.  A
decoder with perfect state information is a mapping
\begin{equation}\label{eq:decoderCSI}
g\colon \set{Y}^n\times \set{S}^n\to\set{M},\quad (y^n,s^n)\mapsto \hat{m}
\end{equation}
with the decoded message now being $g(Y^{n},S^{n})$.

A rate $R$ is said to be achievable if there exists a sequence of
helpers, encoders, and decoders of said rate of average probability of
error tending to zero, where the average is over the uniformly drawn
message and the random state sequence. The capacity is the supremum of
the achievable rates.

We recall that the capacity of an SD-DMC with perfect strictly-causal state information at the encoder equals the capacity without state information; i.e., strictly causal state information does not increase capacity. The capacity with perfect causal state information is given by \cite{shannon58}\begin{subequations}\label{eq:shannon}
\begin{equation}
\max I(U;Y),
\end{equation}
where the maximum is over the choice of the set $\set{U}$ and over the joint distributions of the form
\begin{IEEEeqnarray}{c}
P_S(s) \, P_U(u) \, P_{X|US} (x|u,s) \, W(y|x,s).
\end{IEEEeqnarray}
\end{subequations}
Without reducing the maximum, the conditional PMF $P_{X|US}$ above can be chosen to be deterministic. Thus, $\set{U}$ can be taken as the set of all mappings from $\set{S}$ to $\set{X}$. We sometimes refer to these mappings as ``Shannon strategies.''

\medskip

The rest of this paper is organized as follows:
Section~\ref{sec:capacity} presents our capacity results for various
settings; Section~\ref{sec:counterexamples} provides three
counterexamples demonstrating, respectively, that revealing the
message to the helper increases capacity, that symbol-by-symbol
helpers need not be optimal, and that they need not be optimal even
when the help is provided to both encoder and decoder;
Section~\ref{sec:blockMarkov} presents a block-Markov scheme that can
outperform all symbol-by-symbol helpers; and
Section~\ref{sec:conclusion} offers some intuition
for some of the results and concludes the paper.




\section{Capacity Results}\label{sec:capacity}
\subsection{Symbol-by-symbol helper}
The following result on symbol-by-symbol helpers is a small variation
on known results. Given a fixed symbol-by-symbol helper, we can view
$h(S_{i})$ as a new state and then invoke Shannon's result to obtain
the best achievable rate that can be achieved with said helper
\cite{caireshamai99,jafar06}. We include this theorem
for completeness and because our definition of a symbol-by-symbol
helper does not require that the functions in~\eqref{eq:sbs} be time
invariant, i.e., not depend on~$i$. Also, we allow for the past states
(unquantized) to be revealed to the encoder. The converse must thus be
slightly modified.

\medskip

\begin{theorem}\label{thm:sbs}
  If only symbol-by-symbol helpers are permitted, then the capacity is
\begin{subequations}\label{eq:Cap_sbs}
\begin{equation}
\max I(U;Y),
\end{equation}
where the maximum is over the choice of a set $\set{U}$
and over the joint distributions of the form
\begin{IEEEeqnarray}{c}\label{eq:sbsjoint}
P_S(s) \, P_U(u) \,  P_{T|S} (t|s) \,  P_{X|UT} (x|u,t) \, W(y|x,s),\IEEEeqnarraynumspace
\end{IEEEeqnarray}
\end{subequations}
where---without reducing the maximum---the cardinality of $\set{U}$
may be restricted to $\card{\set{X}}^{\card{\set{T}}}$, and the  conditional PMFs $P_{T|S}$
and $P_{X|UT}$ can be chosen to be deterministic.

This is also the capacity if the time-$i$ channel input produced by
the encoder may depend not only on $m$ and $T^{i}$ but also on the
past states $S^{i-1}$, i.e., 
when the encoder's time-$i$ mapping is of the form
\begin{equation}\label{eq:scenc}
f_i\colon \set{M}\times\set{T}^i\times \set{S}^{i-1}\to \set{X},\quad (m,t^i,s^{i-1}) \mapsto x_i.
\end{equation}
\end{theorem}

\begin{IEEEproof}
  In the direct part, we shall prove that \eqref{eq:Cap_sbs} is
  achievable also in the absence of perfect past state information; in
  the converse part, we shall prove that one cannot achieve a higher
  rate even in its presence. That $P_{T|S}$ and $P_{X|UT}$ can be
  chosen to be deterministic follows because, for any fixed $P_U$, the
  mutual information $I(U;Y)$ is convex in $P_{Y|U}$, and because
  $P_{Y|U}$ is linear in both $P_{T|S}$ and $P_{X|UT}$. Once
  $P_{X|UT}$ is chosen to be deterministic, $\set{U}$ can be
  restricted to comprise the mappings from $\set{T}$ to $\set{X}$,
  hence its cardinality need not exceed $|\set{X}|^{|\set{T}|}$.

\emph{Direct part.} Choose $P_{T|S}$ to be a deterministic mapping
that achieves the maximum in \eqref{eq:Cap_sbs}, so $T=h(S)$ with
probability one for some $h(\cdot)$. The helper produces $t_i=h(s_i)$
for every~$i$. This implies that the sequence $T^n$ is IID. The
encoder and the decoder treat $T$ as the  channel state governing the SD-DMC
\begin{IEEEeqnarray}{rCl}
  \tilde{W}(y|x,t) & = & \sum_{s \in \set{S}} P_{S|T}(s|t) \, W(y|x,s)
\end{IEEEeqnarray}
and can thus achieve \eqref{eq:Cap_sbs} using Shannon strategies
\cite{shannon58}; cf.\ \eqref{eq:shannon}.

\emph{Converse part.} Given mappings $\{f_i\}$ as in 
\eqref{eq:scenc}, we can use Fano's inequality to infer the existence
of some sequence $\{\epsilon_n\}$ tending to zero for which
\begin{IEEEeqnarray}{rCl}
n(R-\epsilon_n) & \le & I(M;Y^n)\\
& = & \sum_{i=1}^n I(M;Y_i|Y^{i-1})\\
& \le & \sum_{i=1}^n I(M,Y^{i-1};Y_i)\\
& \le & \sum_{i=1}^n I(M,Y^{i-1},S^{i-1};Y_i)\\
& = &\sum_{i=1}^n I(U_i;Y_i),
\end{IEEEeqnarray}
where in the last equality we defined, for every $i\in\{1,\ldots,n\}$,
\begin{equation}
U_i \triangleq (M,Y^{i-1},S^{i-1}).
\end{equation}
It remains to check that, for every $i$, the joint distribution of $(S_i,T_i,U_i,X_i,Y_i)$ has the form \eqref{eq:sbsjoint}. To this end, it suffices to verify that the following three conditions are satisfied:
\begin{subequations}
\begin{IEEEeqnarray}{c}
(U_i,T_i) \markov (X_i,S_i) \markov Y_i\label{eq:con1}\\
S_i \markov (U_i,T_i) \markov X_i\label{eq:con2}\\
U_i\indep (S_i,T_i).\label{eq:con3}
\end{IEEEeqnarray}
\end{subequations}
Indeed, \eqref{eq:con1} is satisfied because the channel is memoryless; \eqref{eq:con2} because, given $U_i=(M,Y^{i-1},S^{i-1})$, one can compute $T^{i-1}$ (as a function of $S^{i-1}$), and $X_i$ is determined by $(M,T^i,S^{i-1})$; and \eqref{eq:con3} because the state sequence is IID, and because, with a symbol-by-symbol helper, $T_i$ depends on~$S_i$ alone and not on $S^{i-1}$. This completes the proof of the converse part and hence of the theorem.
\end{IEEEproof}
\medskip

The rate \eqref{eq:Cap_sbs} is a lower bound to the causal-helper
capacity, because every symbol-by-symbol helper is causal. This bound,
however, is not tight:

\medskip

\begin{remark}
For some channels, there exist causal helpers that outperform all
symbol-by-symbol helpers; see Example~\ref{ex:from_hell} ahead.
\end{remark}

\medskip

This bound does, however, characterize the SD-DMCs having positive
causal-helper capacity:

\medskip 

\begin{theorem}\label{thm:positive}
  Subject to~\eqref{eq:assumptionT}, the following statements are
  equivalent:
  \begin{enumerate}
  \item \label{item1}The causal-helper capacity is positive.
  \item \label{item2} The symbol-by-symbol helper capacity is positive.
  \item \label{item3}The capacity of the SD-DMC when the state is revealed
    perfectly to both
    encoder and decoder is positive. Equivalently, there exists some state $s^{\star}
    \in \set{S}$ with $P_{S}(s^{\star}) > 0$ and some $x_1,x_2\in\set{X}$ such that
\begin{equation}
W(\cdot|x_1,s^\star) \neq W(\cdot|x_2,s^\star), \label{eq:Wneq}
\end{equation}    
indicating that the output PMFs induced by $(x_1,s^\star)$ and by $(x_2,s^\star)$ differ.
  \end{enumerate}
\end{theorem}
\begin{IEEEproof}
We first verify the equivalence of the two conditions in~\ref{item3}). 
The capacity of an SD-DMC with perfect state information at both encoder and decoder
  is given by (see \cite{elgamalkim11} and references therein)
\begin{subequations}
\begin{equation}
\max_{P_{X|S}} I(X;Y|S),
\end{equation}
where the conditional mutual information is computed with respect to the joint
PMF
\begin{equation}
P_S(s) \, P_{X|S}(x|s) \, W(y|x,s).
\end{equation}
\end{subequations}
This capacity is positive if, and only if, there exists some $s^\star \in\set{S}$ such that $P_S(s^\star)>0$ and 
\begin{equation}\label{eq:IXYSg0}
\max_{P_{X|S = s^{\star}}} I(X;Y|S=s^\star) >0.
\end{equation}
Inequality~\eqref{eq:IXYSg0} holds if, and only if, there exist $x_1,x_2\in\set{X}$ that satisfy \eqref{eq:Wneq}.

  We next show the equivalence of \ref{item1}), \ref{item2}), and \ref{item3}). Of the three capacities, the symbol-by-symbol helper capacity is
  smallest, and the perfect-encoder-and-decoder-state-information
  capacity is largest. It thus suffices to prove, as we proceed to do, that if
  the latter is positive, then so is the former.

  Assume that~\ref{item3}) holds. By Assumption~\eqref{eq:assumptionT}, $\set{T}$ has at least two
distinct elements. Call them $0$ and $1$.  Now consider the time-invariant
symbol-by-symbol mapping, $h_{i}(\cdot) = h(\cdot)$
where
\begin{IEEEeqnarray}{rCl}
  h(s) & = & \begin{cases}
               0 & \text{if $s = s^{\star}$}\\
               1 & \text{if } s \in \set{S}\setminus \{s^\star\}. \end{cases}
\end{IEEEeqnarray}
The helper thus only tells the encoder whether or not the current
state is $s^{\star}$.
Let $P_U$ be the uniform distribution over the two mappings $u_1,u_2$ from $\set{T}$ to $\set{X}$, where
\begin{subequations}
\begin{IEEEeqnarray}{rCl}
u_1(0) & = & x_1\\
u_2(0) & = & x_2\\
u_1(1)=u_2(1) & = & x_0,
\end{IEEEeqnarray}
\end{subequations}
where $x_0$ can be chosen to be any element of $\set{X}$ (not
necessarily different from $x_1$ or $x_2$). More formally, we choose
$\set{U} = \{1,2\}$; the PMF $P_{U}$ to be uniform over $\set{U}$; and
we choose $P_{X|UT}$ to be deterministic, so $x = x(u,t)$, where
\begin{IEEEeqnarray}{rCl}
  x(u = 1,t) & = & \begin{cases} 
                 x_{1} & \text{if $t = 0$} \\
                 x_{0} & \text{if $t = 1$}
               \end{cases}
\end{IEEEeqnarray}
and
\begin{IEEEeqnarray}{rCl}
  x(u = 2,t) & = & \begin{cases} 
                 x_{2} & \text{if $t = 0$} \\
                 x_{0} & \text{if $t = 1$}.
               \end{cases}
\end{IEEEeqnarray}

This choice of $P_{U}$ and $P_{X|UT}$ implies that, or every $y\in\set{Y}$,
\begin{IEEEeqnarray}{rCl}
P_{Y|U}(y|u_1) & = & P_S(s^\star) \, W(y|x_1,s^\star) + \sum_{s\neq s^\star}
P_S(s) \, W(y|x_0,s)\nonumber\\*[-3mm] \\
P_{Y|U}(y|u_2) & = & P_S(s^\star) \, W(y|x_2,s^\star) + \sum_{s\neq s^\star}
P_S(s) \, W(y|x_0,s).\nonumber\\*[-3mm]
\end{IEEEeqnarray}
By \eqref{eq:Wneq}, 
\begin{equation}
P_{Y|U}(\cdot|u_1) \neq P_{Y|U}(\cdot|u_2),
\end{equation}
which implies that
\begin{equation}
I(U;Y)>0.
\end{equation}
It then follows from Theorem~\ref{thm:sbs} that the capacity of this
channel with a symbol-by-symbol helper is positive.
\end{IEEEproof}



\subsection{Helper cognizant of message}

When a causal helper is cognizant of the message that the encoder
wishes to send, its time-$i$ help is characterized by a mapping of the
form
\begin{equation}\label{eq:mchelper}
h_i\colon \set{M}\times\set{S}^i \to \set{T},\quad (m,s^i)\mapsto t_i.
\end{equation}
Said helper is a \emph{symbol-by-symbol message-cognizant helper} if
this function has the form
\begin{equation}
h_i \colon \set{M}\times \set{S}\to\set{T},\quad (m,s_i) \mapsto t_i.
\end{equation}
In both cases the encoder is as before, i.e., characterized by
mappings of the form~\eqref{eq:encoder_helped_causally}.
For this setting, symbol-by-symbol message-cognizant helpers achieve
capacity:

\medskip

\begin{theorem}\label{thm:hkm}
  The capacity of an SD-DMC with a message-cognizant causal helper is
  achieved by a message-cognizant symbol-by-symbol helper and is given
  by 
\begin{subequations}\label{eq:hkm}
\begin{equation}
\max I(U;Y),
\end{equation}
where the maximum is over the choice of a set $\set{U}$ and over the joint distributions of the form
\begin{IEEEeqnarray}{c}\label{eq:hkmjoint}
P_S(s) \, P_{U}(u) \, P_{T|US} (t|u,s) \, P_{X|UT} (x|u,t) \, W(y|x,s), \IEEEeqnarraynumspace
\end{IEEEeqnarray}
\end{subequations}
where, without loss of optimality, $P_{T|US}$ and $P_{X|UT}$ can be
chosen to be deterministic.
\end{theorem}

\medskip

\begin{IEEEproof}
  That $P_{T|US}$ and $P_{X|UT}$ can be chosen to be deterministic
  follows because, for any fixed $P_U$, $I(U;Y)$ is convex in
  $P_{Y|U}$, and because $P_{Y|U}$ is linear in both $P_{T|US}$ and
  $P_{X|UT}$. (See~\eqref{eq:super_ch_law} ahead.)

  \emph{Direct part.} Fix a PMF $P_U$, a (deterministic) mapping
  $h\colon \set{U}\times\set{S}\to\set{T}$, and a (deterministic)
  mapping $f\colon \set{U}\times\set{T}\to \set{X}$ that achieve the
  maximum in \eqref{eq:hkm}. Consider the ``super channel'' with input
  alphabet~$\set{U}$, output alphabet~$\set{Y}$, and of the
  conditional law
  \begin{IEEEeqnarray}{rCl}
    \tilde{W}(y|u) & = & \sum_{\substack{s \in \set{S}\\t\in\set{T}}} P_{S}(s) \,
    P_{T|US}(t|u,s) \, P_{X|UT}(x|u,t) \, W(y | x,s) \nonumber\\*[-5mm] \label{eq:super_ch_law}
  \end{IEEEeqnarray}
  induced by~\eqref{eq:hkmjoint}.

  We will show that, given any codebook
  $\bigl\{ u^{n}(m) \bigr\}_{m \in \set{M}}$ for this super channel,
  there exists a scheme with a message-dependent symbol-by-symbol
  helper that achieves the same error probability on the original
  channel. To this end, suppose that $m\in\set{M}$ is the message to
  be transmitted, and $u^n(m)$ is the corresponding codeword for 
  the super channel. Since the helper knows the
  message $m$, it also knows $u^{n}(m)$.  In the proposed scheme 
  for the original channel, the helper produces
\begin{equation}
  t_i = h \bigl( u_i(m),s_i \bigr),\qquad i =1,\ldots, n.\IEEEeqnarraynumspace
\end{equation}
The encoder---that knows $u^{n}(m)$ (because it is cognizant of~$m$) and that
obtains $t_{i}$ from the helper---sends
\begin{equation}
x_i = f\bigl(u_i(m),t_i \bigr).
\end{equation}
The conditional distribution $P_{Y^{n}|M}(y^{n}|m)$ of $Y^{n}$ given $m$ that this scheme
induces is
\begin{IEEEeqnarray}{rCl}
  P_{Y^{n}|M}(y^{n}|m) & = & \prod_{i=1}^{n} \tilde{W}\bigl(y_{i} \big| u_{i}(m) \bigr),
\end{IEEEeqnarray}
so the probability of error of this scheme is identical to that of the
code on the super channel.
The proposed scheme can thus achieve the capacity of the super
channel, which equals~\eqref{eq:hkm}.

\emph{Converse part.} Fix mappings $\{h_i\}$ as in \eqref{eq:mchelper}, and define
\begin{equation}
U_i \triangleq (M,T^{i-1},Y^{i-1}),\qquad i=1,\ldots,n. \IEEEeqnarraynumspace
\end{equation}
From Fano's inequality we infer that, if a uniformly drawn message $M$ is
transmitted using a helping scheme with vanishing probability of
error, then, for some $\epsilon_n$ that tends to zero as $n\to\infty$,
\begin{IEEEeqnarray}{rCl}
n(R-\epsilon_n) & \le & I(M;Y^n)\\
& = & \sum_{i=1}^n I(M;Y_i|Y^{i-1})\\
& \le & \sum_{i=1}^n I(M,Y^{i-1};Y_i)\\
& \le & \sum_{i=1}^n I(U_i;Y_i).
\end{IEEEeqnarray}
It remains to check that the joint distribution of
$(S_i,T_i,U_i,X_i,Y_i)$ has the form \eqref{eq:hkmjoint}, i.e., that
the following Markov and independence conditions are satisfied:
\begin{subequations}
\begin{IEEEeqnarray}{c}
(U_i,T_i) \markov (X_i,S_i) \markov Y_i\label{eq:hkmcon1}\\
S_i \markov (U_i,T_i) \markov X_i\label{eq:hkmcon2}\\
U_i\indep S_i.\label{eq:hkmcon3}
\end{IEEEeqnarray}
\end{subequations}
Here, \eqref{eq:hkmcon1} is satisfied because the channel is an SD-DMC;
\eqref{eq:hkmcon2} because $X_i$ can be determined from $(M,T^i)$ (and
hence from $(U_{i},T_{i})$); and \eqref{eq:hkmcon3} because the state sequence is memoryless. 
\end{IEEEproof}

Note that the difference in \eqref{eq:Cap_sbs} from \eqref{eq:hkm} is that $P_{T|US}$ replaces $P_{T|S}$. This can also be seen in the difference between \eqref{eq:con3} and \eqref{eq:hkmcon3}.

Having the helper be cognizant of the transmitted message is
advantageous. The best message-cognizant causal helper can
outperform all message-oblivious causal helpers:

\medskip

\begin{remark}
  The capacity of an SD-DMC with a causal message-cognizant helper can
  exceed that with the best causal message-oblivious helper; see Example~\ref{ex:hkm}.
  \end{remark}
    
\subsection{Channel state known to decoder}

\begin{theorem}\label{thm:psi_at_decoder}
  When the decoder has perfect state information, i.e., when it is of the
  form~\eqref{eq:decoderCSI}, the causal-helper capacity is given by
\begin{subequations}
\begin{equation}
\max I(X;Y|S),
\end{equation}
where the maximum is over all the joint distributions of the form
\begin{equation}\label{eq:dksjoint}
P_S(s) \, P_{T|S}(t|s) \, P_{X|T}(x|t) \, W(y|x,s).
\end{equation} 
\end{subequations}
Moreover, said capacity can be achieved with a symbol-by-symbol helper.
\end{theorem}

\medskip

\begin{IEEEproof}
\emph{Direct part.} We use a symbol-by-symbol helper and apply
Theorem~\ref{thm:sbs}. Since the decoder is cognizant of
the state, the state can be viewed as part of the channel output,
so we can replace $\max I(U;Y)$ in Theorem~\ref{thm:sbs} with
$\max I(U;Y,S)$, which can be simplified as follows:
\begin{IEEEeqnarray}{rCl}
  \max I(U;Y,S) & = & \max I(U;Y|S) \label{eq:IUYS1}\\
  & = & \max I(X;Y|S),\label{eq:IUYS2}
\end{IEEEeqnarray}
where \eqref{eq:IUYS1} holds because $U$ is independent of $S$
in~\eqref{eq:sbsjoint}; and \eqref{eq:IUYS2} holds because, when both
$P_{T|S}$ and $P_{X|UT}$ are chosen to be deterministic, $X$ is a
function of $U$ and $S$, and because $U\markov (X,S)\markov Y$ forms a
Markov chain. Integrating $U$ out reduces the joint PMF
\eqref{eq:sbsjoint} to \eqref{eq:dksjoint}.

\emph{Converse part.} Fix mappings $\{h_i\}$ as in \eqref{eq:h_general},
and define for every $i\in\{1,\ldots,n\}$
\begin{IEEEeqnarray}{rCl}
U_i  & \triangleq & (M,Y^{i-1})\\
V_i  & \triangleq  & S^{i-1}.
\end{IEEEeqnarray}
From Fano's inequality we infer that, if the decoder is cognizant of
the state, and if a uniformly drawn message $M$ is
transmitted using a helping scheme with vanishing probability of
error, then, for some $\epsilon_n$ that tends to zero as $n\to\infty$,
\begin{IEEEeqnarray}{rCl}
n(R-\epsilon_n) & \le & I(M;Y^n, S^n)\\& = & \sum_{i=1}^n I(M;Y_i,S_i|Y^{i-1},S^{i-1})\\
& \le & \sum_{i=1}^n I(M,Y^{i-1};Y_i,S_i|S^{i-1})\\
& = & \sum_{i=1}^n I(M,Y^{i-1};Y_i|S_i,S^{i-1}) \label{eq:why} \\
& = & \sum_{i=1}^n I(U_i;Y_i|S_i,V_i).
\end{IEEEeqnarray}
The capacity is thus upper-bounded by the maximum of
\begin{equation}\label{eq:IUYSV}
I(U;Y|S,V)
\end{equation}
over the auxilliary sets $\set{U}$ and $\set{V}$, and over the joint distributions of the form
\begin{IEEEeqnarray}{l}
P_V(v) \, P_S(s) \, P_{U|V}(u|v)  \, P_{T|SV}(t|s,v) \nonumber\\*
\qquad\qquad\qquad \cdot P_{X|TUV}(x|t,u,v) \, W(y|x,s). \IEEEeqnarraynumspace
\end{IEEEeqnarray}
We further upper-bound \eqref{eq:IUYSV} by its maximum over $V=v$:
\begin{equation}
I(U;Y|S,V)\le \max_{v\in\set{V}} I(U;Y|S,V=v).
\end{equation}
Since $V\indep S$, we can remove $V$ by fixing $V=v$ 
that achieves the above maximum, so the upper
bound becomes $\max I(U;Y|S)$ over the joint distribution
\eqref{eq:sbsjoint}. As in the direct part, $I(U;Y|S) = I(X;Y|S)$,
which completes the proof.
\end{IEEEproof}

\section{Counterexamples}\label{sec:counterexamples}
\subsection{Revealing the message to the helper increases capacity}
\begin{example}\label{ex:hkm}
The channel input $X$ is a binary tuple
\begin{equation}
X=(A,B)
\end{equation}
with $A,B$ both taking values in $\{0,1\}$. The state $S$ too is a
binary
tuple 
\begin{equation}
S=(S^{(0)},S^{(1)}),
\end{equation}
where $S^{(0)}$ and $S^{(1)}$ are independent and both uniform over $\{0,1\}$. The channel output is
\begin{equation}
Y = (A, B\oplus S^{(A)}).
\end{equation}
The helper's description rate is $1$ bit:
\begin{equation}
\set{T}=\{0,1\},
\end{equation}
so it cannot fully describe the state.
\end{example}

\medskip

\begin{claim}
For the SD-DMC of Example~\ref{ex:hkm}:
\begin{enumerate}
\item The capacity with a message-cognizant causal helper is $2$ bits
  per channel use.
\item The capacity with a message-oblivious causal helper is $\log 3$,
  which can be achieved with a symbol-by-symbol helper. Furthermore,
  $\log 3$ is the capacity also when the message-oblivious helper is
  noncausal and the help $T$ is provided also to the decoder.
\end{enumerate}
\end{claim}

\medskip

\begin{IEEEproof}
  \emph{Message-cognizant helper.} The capacity cannot exceed
  $\log|\set{X}| = \log |\set{Y}|= 2 $ bits, so we focus on
  achievability and describe a scheme that
  can convey $2$ bits error-free in a single channel use. Let
  $\alpha$ and $\beta$ denote the information bits to be
  conveyed. Since the helper is cognizant not only of the state but
  also of $(\alpha,\beta)$, it can 
  assist the encoder by providing it with
\begin{equation}
T = S^{(\alpha)}.
\end{equation}
The encoder can then produce the channel input
\begin{equation}
X=(\alpha, \beta \oplus T) = (\alpha,\beta\oplus S^{(\alpha)}),
\end{equation}
where $\oplus$ denotes modular-$2$ addition. The output is then
\begin{equation}
Y = (\alpha,\beta\oplus S^{(\alpha)}\oplus S^{(\alpha)}) =
(\alpha,\beta) \label{eq:Bibi100}
\end{equation}
and both bits are correctly conveyed without the need for
decoding. (The achievability of $2$ bits could also be deduced from
Theorem~\ref{thm:hkm} by choosing $U = (A, V)$ uniform on
$\{0,1\}\times\{0,1\}$, $T = S^{(A)}$, and $ X = (A, V\oplus T)$.)

\medskip

We next turn to the message-oblivious helper. But first we
recall a result on the ``sum channel''
\cite[Problem~4.18]{gallager68}, \cite[Problem~7.28]{coverthomas06}.

\textit{Sum channel.} Consider a discrete memoryless channel that is the ``sum'' of $\ell$ disjoint sub-channels in the following sense: 
\begin{IEEEeqnarray}{ll}
\set{X} = \set{X}_1\cup \cdots \cup \set{X}_\ell,\quad &\set{X}_i \cap \set{X}_j = \emptyset, i\neq j \IEEEeqnarraynumspace \\
\set{Y} = \set{Y}_1\cup \cdots \cup \set{Y}_\ell,\quad &\set{Y}_i \cap \set{Y}_j = \emptyset, i\neq j\\
\Pr(Y\in\set{Y}_i|X=x)  =  1,\quad &x\in\set{X}_i.
\end{IEEEeqnarray}
Let $C_i$ denote the capacity of the $i$-th sub-channel, i.e., the channel with input alphabet $\set{X}_i$ and output alphabet $\set{Y}_i$. Then the capacity of the sum channel is 
\begin{equation}
C = \log \sum_{i=1}^\ell 2^{C_i}.
\end{equation}

\textit{Message-oblivious helper, direct part.}
Let the (symbol-by-symbol) helper produce
\begin{equation}
T=S^{(0)}.
\end{equation}
Since the first component of the output $Y$ is equal to the first
component of the input $X$ (see \eqref{eq:Bibi100}), we can view the channel
as the sum of two channels: the first where $A=0$ so
$\set{X}_{1} = \set{Y}_{1} = \{(0,0), (0,1)\}$ 
and the second where $A = 1$ so $\set{X}_{2} = \set{Y}_{2} = \{(1,0),
(1,1)\}$.

The encoders of both channels observe $T$, but the second ignores
it. The first encoder, cognizant of $S^{(A)}=S^{(0)}$, can perfectly
control the second output bit $B\oplus S^{(0)}$, so $C_1=1$ bit.
The encoder for the second sub-channel, where $A=1$, is incognizant of
$S^{(A)}=S^{(1)}$, so the sub-channel's output bit $B\oplus S^{(1)}$
is random and independent of the input, so $C_2=0$. The capacity of
the sum channel is thus
\begin{equation}
\log \left(2^1 + 2^0\right) = \log 3.
\end{equation}

\textit{Message-oblivious helper, converse part.} We first consider
the special case where the helper is symbol-by-symbol. This step is
unnecessary but sheds light on the proof of the general case. Assuming
that the help is provided also to the decoder and treating $T=h(S)$ as
the channel meta state, we can express the capacity as
\begin{equation}
\max I(X;Y|T),
\end{equation}
where the maximization is over the conditional law $P_{X|T}$. When we
fix any $T=t$, the channel becomes a sum channel. The first
sub-channel is where $A=0$. With $A=0$ fixed, the maximum value of
$H(Y|T=t)$ is $1$ bit, which is achieved when $B$ is uniform, whereas
\begin{equation}
H(Y|X,T=t) = H(B\oplus S^{(0)}|B,T=t) = H(S^{(0)}|T=t).
\end{equation}
We conclude that, conditional on $T=t$,  the capacity of the first sub-channel is
\begin{equation}
C_1 = 1-H(S^{(0)}|T=t).
\end{equation}
Similarly, under the same conditioning, the capacity of the second sub-channel is
\begin{equation}
C_2 = 1-H(S^{(1)}|T=t).
\end{equation}
Conditional on $T=t$, the capacity of the sum channel is
\begin{IEEEeqnarray}{rCl}
\IEEEeqnarraymulticol{3}{l}{\max I(X;Y|T=t)}\nonumber\\*
\qquad& = & \log\left( 2^{1-H(S^{(0)}|T=t)} + 2^{1-H(S^{(1)}|T=t)}\right).
\end{IEEEeqnarray}
This and the inequality
\begin{equation}\label{eq:21a}
2^{1-a} \le 2-a,\qquad a\in[0,1], 
\end{equation}
imply that
\begin{IEEEeqnarray}{rCl}
\IEEEeqnarraymulticol{3}{l}{
\max I(X;Y|T=t)
}\nonumber\\* \,\,
& \le & \log \Bigl( 2 - H(S^{(0)}|T=t) + 2 - H(S^{(1)}|T=t) \Bigr) \IEEEeqnarraynumspace \\
& \le & \log \Bigl( 4 - H(S^{(0)},S^{(1)}|T=t) \Bigr).
\end{IEEEeqnarray}
Averaging over $t$ and employing Jensen's inequality, we obtain an upper
bound on the capacity with a message-oblivious symbol-by-symbol helper
when the help is provided also to the decoder:
\begin{IEEEeqnarray}{rCl}
C & \le & \max \sum_t P_T(t) \log \left(4 - H(S^{(0)},S^{(1)}|T=t)\right) \IEEEeqnarraynumspace \\
& \le & \max \log \left( 4 - H(S^{(0)},S^{(1)}|T)\right) \label{eq:C72}\\
& \le & \log (4-1) \label{eq:C73}\\
& = & \log 3,
\end{IEEEeqnarray}
where \eqref{eq:C72} follows because $\log$ is concave; and \eqref{eq:C73} because
\begin{IEEEeqnarray}{rCl}
H(S^{(0)},S^{(1)}|T) & = & \underbrace{H(S^{(0)},S^{(1)})}_{=2} - \underbrace{I(S^{(0)},S^{(1)};T)}_{\le H(T)\le 1} \IEEEeqnarraynumspace \\
& \ge & 1.
\end{IEEEeqnarray}

We next show that one cannot achieve any rate larger than $\log 3$
even with a noncausal helper, and when the help is provided also to the
decoder. In the rest of this proof, we
shall slightly abuse notation to use $T$ to denote the $n$-letter
assistance, i.e., $T$ is a function of $S^n$ and takes values in
$\{0,1\}^n$. Using Fano's inequality we can infer that, if there
exists a coding scheme of rate $R$ that has vanishing error
probability, then, for some $\epsilon_n$ that tends to zero as
$n\to\infty$,
\begin{equation}
n(R-\epsilon_n) \le I(X^n;Y^n|T),
\end{equation}
so we shall bound the maximum of $I(X^n;Y^n|T)$ over all
$P_{X^{n}|T}$.  Given $T=t$, the $n$-letter channel can be viewed as a
sum channel containing $2^n$ sub-channels, each corresponding to a
different choice of the binary $n$-tuple
$(A_1,\ldots,A_n)$. Conditional on $T=t$, the
capacity of the sub-channel corresponding to $A_1=a_1,\ldots,A_n=a_n$ is
\begin{equation}
\exp_2 \left\{ n - H\left(S_1^{(a_1)},\ldots,S_n^{(a_n)}\middle| T=t  \right) \right\},
\end{equation}
where $\exp_2\{\xi\}$ denotes $2^{\xi}$. 
Hence, 
\begin{IEEEeqnarray}{rCl}
\IEEEeqnarraymulticol{3}{l}{
  \max_{P_{X^n|T=t}}
  I(X^n;Y^n|T=t)
  }\nonumber\\*
   & = & \log \sum_{a^n\in\{0,1\}^n} \exp_2 \left\{n - H\left(S_1^{(a_1)},\ldots,S_n^{(a_n)}\middle| T=t  \right)\right\}\nonumber\\*[-3mm]\\
& = & \log \sum_{a^n\in\{0,1\}^n} \prod_{i=1}^n \exp_2  \nonumber\\*
& & \qquad\quad\left\{ 1 - H\left(S_i^{(a_i)}\middle |S_1^{(a_1)},\ldots,S_{i-1}^{(a_{i-1})}, T=t\right)\right\} \\
& \le  & \log \sum_{a^n\in\{0,1\}^n} \prod_{i=1}^n \exp_2 \left\{ 1 - H\left(S_i^{(a_i)}\middle |S^{i-1}, T=t) \right)\right\}\IEEEeqnarraynumspace\\
& = & \log \prod_{i=1}^n \Bigg( \exp_2 \left\{ 1 - H\left(S_i^{(0)}\middle| S^{i-1}, T=t\right)\right\} \nonumber\\*
& & \qquad\qquad{}+ \exp_2 \left\{ 1 - H\left(S_i^{(1)}\middle| S^{i-1}, T=t\right)\right\} \Bigg) \\
& = & \sum_{i=1}^n \log \Bigg( \exp_2 \left\{ 1 - H\left(S_i^{(0)}\middle| S^{i-1}, T=t\right)\right\} \nonumber\\*
& & \qquad\qquad{}+ \exp_2 \left\{ 1 - H\left(S_i^{(1)}\middle| S^{i-1}, T=t\right)\right\} \Bigg)\\
& \le & \sum_{i=1}^n \log \bigg( 2 - H\left(S_i^{(0)}\middle| S^{i-1}, T=t\right) \nonumber\\*
& & \qquad\qquad{} + 2 - H\left(S_i^{(1)}\middle| S^{i-1}, T=t\right)\bigg) \label{eq:90}\\
& \le &  \sum_{i=1}^n \log \bigg( 4 - H\left(S_i\middle|S^{i-1},T=t\right) \bigg)\\
& \le & n \log \left( 4 - \frac{1}{n} \sum_{i=1}^n H\left(S_i\middle| S^{i-1},T=t\right) \right) \label{eq:85}\\
& = & n \log \left( 4 - \frac{H(S^n|T=t)}{n}\right), \label{eq:92}
\end{IEEEeqnarray}
where \eqref{eq:90} follows from \eqref{eq:21a}, and \eqref{eq:85}
follows from the concavity of the logarithm.
Averaging \eqref{eq:92} over $T$ and again using the concavity of the
logarithm, we obtain the bound
\begin{IEEEeqnarray}{rCl}
I(X^n;Y^n|T) & \le & n \log \left( 4 - \frac{H(S^n|T)}{n}\right). \label{eq:87}
\end{IEEEeqnarray}
Note that
\begin{IEEEeqnarray}{rCl}
H(S^n|T) & = & H(S^n) - I(S^n;T)\\
& \ge & H(S^n) - H(T)\\
& \ge & 2n - n\\
& = & n.
\end{IEEEeqnarray}
This and \eqref{eq:87} imply that
\begin{equation}
I(X^n;Y^n|T) \le n \log 3,
\end{equation}
which completes the proof.
\end{IEEEproof}

\subsection{A causal helper that outperforms all symbol-by-symbol helpers}
\begin{example}\label{ex:from_hell}
The channel input is
\begin{equation}
X=(A,B,C),
\end{equation}
where $A,B$ take values in $\{0,1\}$, and $C$ in $\{0,1\}^\eta$ for
some integer $\eta$ (larger than $10$). The state is a
pair
\begin{equation}
S=(S^{(0)},S^{(1)}),
\end{equation}
where $S^{(0)}$ and $S^{(1)}$ are independent, both uniform on $\{0,1\}$. The output is
\begin{equation}
Y = (A', D^{(0)},D^{(1)}),
\end{equation}
with $A'$ taking values in $\{0,1\}$, and with $D^{(0)}$ and $D^{(1)}$ in $\{0,1\}^\eta$.

The channel law is the following:
\begin{itemize}
\item Conditional on $X$ and $S$ with $B \neq S^{(A)}$, the output $Y$
  is uniformly distributed over its alphabet.
\item Conditional on $X$ and $S$ with $B = S^{(A)}$, 
\begin{subequations}\label{eq:goodlaw}
\begin{IEEEeqnarray}{rCl}
A' & = & A \label{eq:Jon50} \\ 
D^{(B)} & = & C
\end{IEEEeqnarray}
deterministically, and 
\begin{IEEEeqnarray}{rCl}
D^{(B\oplus 1)} & \sim & \textnormal{equiprobable over $\{0,1\}^\eta$}.
\end{IEEEeqnarray}
\end{subequations}
\end{itemize}

The (message-oblivious) helper's description rate is $1$ bit:
\begin{equation}
\set{T}=\{0,1\}.
\end{equation}
\end{example}

\begin{claim}\label{cl:from_hell}
In Example~\ref{ex:from_hell}:
\begin{enumerate}
\item There exists a (non symbol-by-symbol) causal helper 
  that allows for the reliable transmission of $\eta$ bits
  per channel use.
\item When restricted to symbol-by-symbol helpers, the capacity is
  strictly less than $\eta$. 
\end{enumerate}
\end{claim}

\begin{IEEEproof}
  \emph{General causal helper.} To prove the achievability of $\eta$
  bits per channel use with a causal (non symbol-by-symbol) helper,
  consider the following coding scheme. Represent the message that is
  to be transmitted in $n$ channel uses as a
  sequence $\alpha_1,\alpha_2,\ldots, \alpha_{n-1}$ of binary
  $\eta$-tuples, so $\alpha_i\in\{0,1\}^\eta$. Set $\alpha_{n}$ to be
  some arbitrary $\eta$-tuple, say all-zero. The transmission rate is
  thus $(n-1) \eta / n$, which approaches $\eta$ as $n \to \infty$.

  Define
\begin{equation}
T_0 \triangleq 0.
\end{equation}
At each time $i$, the help is 
\begin{IEEEeqnarray}{rCl}
T_i & = & S_i^{(T_{i-1})},
\end{IEEEeqnarray}
and the channel input is
\begin{IEEEeqnarray}{rCl}
A_i & = & T_{i-1} \label{eq:Jon80} \\ 
B_i & = & T_i\\
C_i & = & \alpha_i.
\end{IEEEeqnarray}
This guarantees that, at each $i$, 
\begin{equation}
B_i = S^{(A_i)},
\end{equation}
so the channel behaves according to
\eqref{eq:goodlaw}. Moreover, \eqref{eq:Jon80} and
\eqref{eq:Jon50} imply that from $Y_{i+1}$---specifically from
$A'_{i+1}$---the decoder will learn
$B_i$ because $B_i=A_{i+1}=A_{i+1}'$. It will then be able to recover
$\alpha_i$ without error by reading $D_i^{(B_i)}$ (which was received
at time $i$). In this way, 
the decoder
recovers $\alpha_{1}, \ldots, \alpha_{n-1}$ error free.
This concludes the proof of the first part of the claim.

\medskip

\emph{Symbol-by-symbol helper.} There are ostensibly $2^{4}$
symbol-by-symbol helpers. But $T$ and $T \oplus 1$ give identical
performance. Likewise swapping $S^{(0)}$ and $S^{(1)}$ or replacing
either (or both of them) with the complement does not change
performance. After accounting for these symmetries, we must only
analyze three symbol-by-symbol helpers:
\begin{IEEEeqnarray}{rCl}
T & = & S^{(0)}\\
T & = & S^{(0)} \land S^{(1)}\\
T & = & S^{(0)} \oplus S^{(1)}.
\end{IEEEeqnarray}

To analyze $T = S^{(0)} \oplus S^{(1)}$, define $E = E(X,S)$ as
\begin{IEEEeqnarray}{rCl}
  E = \begin{cases}\label{eq:defE}
        1, & \text{if $B = S^{(A)}$} \\
        0, & \text{otherwise}.
        \end{cases}
\end{IEEEeqnarray}
Recall that, conditional on $(X,S)$ with $E(X,S)$ being zero, the output
$Y$ is equiprobable over its alphabet. We can upper-bound the capacity
by assuming that $T$ is revealed also to the decoder and then upper-bounding
$\max I(X;Y|T)$. This we do as follows:
\begin{IEEEeqnarray}{rCl}
  I(X;Y|T) & \leq & I(X;Y,E|T) \\
  & \leq & 1 + I(X;Y|E,T) \\
  & = & 1 + \sum_{t} P_{T}(t) \, I(X;Y|E,T=t). \IEEEeqnarraynumspace
\end{IEEEeqnarray}
For each $t \in \{0,1\}$
\begin{IEEEeqnarray}{rCl}
  I(X;Y|E,T=t) & = & P_{E|T}(1|t) \, I(X;Y|E=1, T=t) \IEEEeqnarraynumspace \\
  & \leq & P_{E|T}(1|t) \, \log |\set{X}| \\
  & = & \frac{1}{2} \, (\eta + 2),
\end{IEEEeqnarray}
where the first equality holds because $I(X;Y|E=0, T=t)$ is zero; and the last equality because, irrespectively of $P_{X|T}$, the probability of $E(X,S)$ being one is always $1/2$. The
capacity with this symbol-by-symbol helper is thus upper-bounded by $2
+ \eta/2$, which is strictly smaller than $\eta$ whenever $\eta$
exceeds $4$.

Consider now $T = S^{(0)} \land S^{(1)}$. Upper-bounding
the mutual information conditional on $T=1$ by $\log |\set{X}|$,
\begin{IEEEeqnarray}{rCl}
  I(X;Y|T) & \leq & P_{T}(1) \, (\eta + 2) + P_{T}(0) \,
  I(X;Y|T=0)\IEEEeqnarraynumspace \\
  & = & \frac{1}{4} \, (\eta + 2) + \frac{3}{4} \, I(X;Y|T=0).
\end{IEEEeqnarray}
The mutual information term can then be bounded by (with~$E$ defined as in \eqref{eq:defE})
\begin{IEEEeqnarray}{rCl}
  I(X;Y|T=0) & \leq & I(X;Y,E|T=0) \\
  & \leq & 1 + I(X;Y|E,T=0) \\
  & = & 1 + P_{E|T}(1|0) \, (\eta + 2) \\
  & \leq & 1 + \frac{2}{3} \, (\eta + 2),
\end{IEEEeqnarray}
where the last inequality holds because, conditional on $S^{(0)} \land S^{(1)} = 0$, the states
$(0,0), (0,1), (1,0)$ are each of probability $1/3$, so $P_{E|T}(1|0)$ cannot exceed $2/3$. The above
inequalities demonstrate that the capacity with this helper is
upper-bounded by $9/4 + 3\eta/4$. This is strictly smaller than
$\eta$ whenever $\eta \geq 10$.

The final symbol-by-symbol helper $T = S^{(0)}$ can be analyzed using
the duality-based upper bound
\cite[Thm.~5.1]{lapidothmoser03_3}; see the appendix.
\end{IEEEproof}

\subsection{Help to both encoder and decoder}

The following example shows that symbol-by-symbol helpers need not be
optimal even when the help is provided to
both encoder and decoder.

\begin{example}\label{ex:both}
The channel input is binary, and the state is quaternary
\begin{IEEEeqnarray}{rCl}
\set{X} & = & \{0,1\}\\
\set{S} & = &\{0,1,2,3\}.
\end{IEEEeqnarray}
The output is a binary 4-tuple
\begin{equation}
Y = (Y^{(0)}, Y^{(1)},Y^{(2)}, Y^{(3)}) \in \{0,1\}^{4}.
\end{equation}
Conditional on the input $X=x$ and the state $S=s$, the component of
$Y$ that is indexed by~$s$ equals $x$ deterministically
\begin{subequations}
  \begin{IEEEeqnarray}{rCl}
  Y^{(s)} & = & x 
\end{IEEEeqnarray}
and the other components are IID Bernoulli $(1/2)$
\begin{IEEEeqnarray}{rCl}
  \{Y^{(s')}\}_{s' \in \set{S} \setminus \{s\}} & \sim & \textnormal{IID Bern($1/2$)}.
\end{IEEEeqnarray}  
\end{subequations}
The helper's description rate is $1$ bit:
\begin{equation}
\set{T} = \{0,1\}.
\end{equation}
\end{example}

\medskip

\begin{claim}
  If the help in Example~\ref{ex:both} is provided to the encoder
  causally and also to the decoder, then:
\begin{enumerate}
\item The highest capacity achievable with a symbol-by-symbol helper
  is $0.5$ bit.
\item With some causal non symbol-by-symbol helper the capacity is at
  least $0.5+0.1875\log 1.5 \approx 0.61$ bit.
\end{enumerate}
\end{claim}

\begin{IEEEproof} \emph{Symbol-by-symbol helper.} After accounting for
  symmetries and relabelings, only two different symbol-by-symbol
  helpers remain. The first is
\begin{IEEEeqnarray}{rCl}
T & = & \begin{cases} 0, & \text{if $S \in \{0,1\}$}\\ 1,&\textnormal{otherwise.} \end{cases} \label{eq:TS41}
\end{IEEEeqnarray}
To analyze it, suppose that both encoder and decoder know that
$T=0$. The output components $Y^{(2)}$ and $Y^{(3)}$ can then be
discarded because they are known to be independent of the
input. Conditional on $X=x$, the remaining components have the following
distribution:
\begin{equation}\label{eq:Y0Y1}
(Y^{(0)},Y^{(1)}) = \begin{cases} (x,x)& \textnormal{w.p. } 0.5\\ (x,x\oplus 1)& \textnormal{w.p. } 0.25\\ (x\oplus 1,x) & \textnormal{w.p. } 0.25.\end{cases}
\end{equation}
It then follows that
\begin{equation}\label{eq:HY0Y11}
H\left(Y^{(0)},Y^{(1)}\middle|X, T=0\right) = 1.5.
\end{equation}
Since
\begin{equation}\label{eq:HY0Y12}
\max H\left(Y^{(0)},Y^{(1)}\middle| T=0 \right) = 2,
\end{equation}
with the maximum achieved by a uniform $X$,
\begin{IEEEeqnarray}{rCl}
  \max I(X; Y| T=0) & = & 0.5.
\end{IEEEeqnarray}
Both \eqref{eq:HY0Y11} and
\eqref{eq:HY0Y12} continue to hold when, rather than $T=0$, we
consider $T=1$. We thus conclude that the maximum achievable rate with
the helper \eqref{eq:TS41} is $0.5$ bit.

The second symbol-by-symbol helper to be considered is
\begin{IEEEeqnarray}{rCl}
T & = & \begin{cases} 0, & S=0 \\ 1,&\textnormal{otherwise.} \end{cases} \label{eq:TS42}
\end{IEEEeqnarray}
When $T=0$, which happens with probability $0.25$, the decoder knows
that $Y^{(0)}=X$, 
so 
\begin{IEEEeqnarray}{rCl}
  \label{eq:IXY0}
  \max I\bigl(X;Y^{(0)}\bigm|T=0\bigr) & = & 1.
\end{IEEEeqnarray}
When $T=1$, which happens with probability $0.75$, $Y^{(0)}$ is
independent of the input and can be discarded, whereas the conditional
probabilities of the remaining components given $X=x$ can be written
down explicitly as in~\eqref{eq:Y0Y1} (details omitted). From these
probabilities we can compute
\begin{equation}\label{eq:IXY123}
\max I\left(X; Y^{(1)},Y^{(2)},Y^{(3)} \middle| T=1\right) = 1.5-0.75\log 3.
\end{equation}
Using \eqref{eq:IXY0} and \eqref{eq:IXY123}, we obtain that the
maximum rate achievable by the helper \eqref{eq:TS42} is
\begin{equation}
0.25 \cdot 1 + 0.75 \cdot (1.5 - 0.75\log 3) \approx 0.483,
\end{equation}
which is inferior to the capacity with the first helper.  The capacity
with the best symbol-by-symbol helper is thus $0.5$ bit and is
achieved by the helper \eqref{eq:TS41}.

\emph{General helper.} Consider a two-letter helper: over two channel
uses, the helper uses $T_1,T_2$ to describe $S_1$ perfectly and
$S_2$ not at all. (More formally, the time-$(2i+1)$ help
$T_{2i+1}$ and the time-$(2i+2)$ help $T_{2i+2}$ describe the
time-$(2i+1)$ state $S_{2i+1}$ ignoring the time-$(2i+2)$ state
$S_{2i+2}$.) At time $2i+2$ the receiver is cognizant of $S_{2i+1}$
and can hence recover $X_{2i+1}$ (which equals the $S_{2i+1}$-th
component of $Y_{2i+1}$). The bit $X_{2i+1}$ is thus recovered error free.
As for $X_{2i+2}$, it is transmitted with neither encoder nor decoder
cognizant of the state,
hence we can write the probabilities of the different
values of $Y_{2i+2}$ as in~\eqref{eq:Y0Y1} (with the details again
omitted). The channel from $X_{2i+2}$ to $Y_{2i+2}$ is of capacity
\begin{equation}
0.375 \log 1.5.
\end{equation}
We can thus transmit $(1+0.375 \log 1.5)$ bits with two channel uses,
from which the second part of the claim follows.
\end{IEEEproof}

\section{A Block-Markov Scheme}\label{sec:blockMarkov}
\newcommand{\bseq}[2]{\mathbf{#1}^{(#2)}}

Inspired by Example~\ref{ex:from_hell}, we propose a communication
scheme employing block-Markov encoding and backward decoding.  Choose
three finite auxiliary sets $\set{Z}$, $\set{U}$, and $\set{V}$, and
fix a joint distribution of the form
\begin{IEEEeqnarray}{l}
P_S(s) \, P_Z(z) \, P_{T|SZ}(t|s,z) \, P_{U|Z}(u|z) \nonumber\\*
\qquad\qquad\qquad \cdot P_{X|UT}(x|u,t) P_{V|T}(v|t) \, W(y|x,s), \IEEEeqnarraynumspace \label{eq:BMjoint}
\end{IEEEeqnarray}
where $P_{T|SZ}$ and $P_{X|UT}$ are deterministic, so there exist
mappings $f\colon \set{U}\times\set{T} \to \set{X}$ and
$h\colon \set{S}\times\set{Z} \to \set{T}$ such that, with probability
one,
\begin{IEEEeqnarray}{rCl}
T & = & h(S,Z)\\
X & = & f(U,T).
\end{IEEEeqnarray}

In the following, we use boldface letters such as $\vect{x}$ and
$\vect{t}$ to denote length-$n$ vectors, and 
we extend $h$ and $f$ to apply to $n$-length vectors
component-wise. For example, $\vect{t} = h(\vect{s},\vect{z})$
indicates that each component of $\vect{t}$ is the result of applying
$h$ to the corresponding components of $\vect{s}$ and
$\vect{z}$. Component-wise extension of conditional probabilities to
length-$n$ vectors are denoted using the superscript $\times n$ as
in $P_{U|Z}^{\times n} (\vect{u}|\vect{z})$.

The block-Markov scheme we propose divides the transmission time into
$\lambda$ blocks, with each of the first $\lambda-1$ blocks being of
length $n$, and with the last being possibly longer. Since $\lambda$
will be very large, the last block will have negligible effect on the
transmission rate. The scheme employs binning and superposition
coding. In the superposition coding, the Block~$j$ cloud center
$\bseq{z}{j}$---being determined at the end of Block~$(j-1)$---is
known to both encoder and helper before the block begins. The
satellite $\bseq{u}{j}$ is determined by the message~$m_j$ that is
transmitted in Block~$j$. The assistance produced in Block~$j$ is
$\bseq{t}{j}=h(\bseq{s}{j},\bseq{z}{j})$. A $v$-sequence,
$\bseq{v}{j}$, is then chosen based on $\bseq{t}{j}$, and is binned as
in Wyner-Ziv coding \cite{wynerziv76}, treating the outputs
$\bseq{y}{j}$ as side information that is available to the
decoder. The bin index determines the cloud center in Block~$(j+1)$. 
We elaborate below.

Fix three positive rates $R$, $R_v$, and $\tilde{R}$ that will be
specified later. In the following, ``typical'' is short for strongly
typical with respect to corresponding marginal of the joint
distribution \eqref{eq:BMjoint} \cite{csiszarkorner11}. We shall not
explicitly write ``$\epsilon$-typical,'' but it shall be understood
that the implicit parameter $\epsilon$ does not depend on $n$ and can
be chosen arbitrarily close to zero. However, the choice of $\epsilon$
may need to be different in each case below, a technicality that, to
simplify the exposition, we ignore. Furthermore, the bounds we shall
derive on the rates should involve slacks that are related to
$\epsilon$. This too we ignore.

\emph{Codebook construction.}  Generate $2^{n(R_v + \tilde{R})}$ length-$n$ sequences IID $\sim P_V$
\begin{equation} \label{eq:vgen}
\vect{v} (\ell,k),\quad \ell\in\{1,\ldots,2^{n R_v}\},k\in\{1,\ldots,2^{n \tilde{R}}\}.
\end{equation}
Independently of the above, generate $2^{n R_v}$ sequences IID $\sim P_Z$
\begin{equation}
\vect{z} (\ell),\quad \ell \in \{1,\dots,2^{n R_v}\}.
\end{equation}
For each $\ell\in\{1,\dots,2^{n R_v}\}$, independently generate
$2^{n R}$ sequences
$\sim P_{U|Z}^{\times n} \big(\cdot \big|\vect{z}(\ell)\big)$
\begin{equation}
\vect{u} (\ell, m),\quad m \in\{1,\ldots,2^{n R}\}.
\end{equation}

\emph{Encoding.} Let $\ell_0\triangleq 1$. When encoding for Block
$j$, where $j\in\{1,\ldots,\lambda-1\}$, the encoder and the helper
have already identified $\ell_{j-1}$ (which they compute at the end of
Block~$(j-1)$).  They pick the cloud center
\begin{equation}
\bseq{z}{j} = \vect{z}(\ell_{j-1}).
\end{equation}
The helper produces the help
\begin{equation} \label{eq:tjsz}
\bseq{t}{j} = h(\bseq{s}{j},\bseq{z}{j}).
\end{equation}
Denoting the message to be sent over all the blocks
$(m_1,\ldots,m_{\lambda-1})$, the encoder selects the satellite according to the message
$m_j\in\{1,\ldots,2^{nR}\}$:
\begin{equation}
\bseq{u}{j} = \vect{u}(\ell_{j-1},m_j).
\end{equation}
It sends
\begin{equation}\label{eq:xjut}
\bseq{x}{j} = f(\bseq{u}{j},\bseq{t}{j}).
\end{equation}
(Note that the mappings \eqref{eq:tjsz} and \eqref{eq:xjut} are indeed causal.)
The helper and the encoder look for the first pair of indices $(\ell_j,k_j)\in\{1,\ldots,2^{nR_v}\}\times \{1,\ldots,2^{n\tilde{R}}\}$ such that 
\begin{equation}
\bseq{v}{j} = \vect{v}(\ell_j,k_j)
\end{equation}
satisfies
\begin{equation}
\left( \bseq{v}{j}, \bseq{t}{j}\right) \textnormal{ are jointly typical.}\label{eq:vsjt}
\end{equation}
They discard $k_j$ and use $\ell_j$ to pick the cloud center in Block
$(j+1)$. If no such pair of indices can be found, then they choose
$\ell_j=1$. We think of $\ell_{j}$ as a bin index.

In the last block, Block $\lambda$, no message bits are sent, and we
only convey the index $\ell_{\lambda-1}$. To this end, we employ a
symbol-by-symbol helper of positive capacity. For now, we assume that
such a helper exists, and proceed to derive an achievable rate by
analyzing Blocks $1$ through $(\lambda-1)$. Later we will assert that this
assumption is unnecessary, because it is satisfied whenever the rate
whose achievability we seek to prove using the block-Markov
scheme---namely, \eqref{eq:amos_bm_is_good} ahead---is positive.  To
prove the assertion, we shall recall that, by
Theorem~\ref{thm:positive}, a symbol-by-symbol helper of positive
capacity exists whenever the capacity with perfect state information
at both encoder and decoder is positive. To prove the assertion, we
shall thus only need to show that this latter capacity is positive
whenever~\eqref{eq:amos_bm_is_good} is positive. This will follow once
we show that said capacity is
greater than or equal to~\eqref{eq:amos_bm_is_good}.

\emph{Decoding.} Based on the output sequence $\bseq{y}{\lambda}$
received in the last block, Block~$\lambda$, the decoder recovers
$\ell_{\lambda-1}$. It then proceeds with backward decoding from
Block~$(\lambda-1)$ to Block~$1$.  By the time the decoder decodes
Block $j$, it will have already recovered $\ell_j$ (from its decoding
Block $(j+1)$). Denote the recovered value by $\hat{\ell}_j$. To decode
Block~$j$, the decoder looks for indices $\hat{k}_j$,
$\hat{\ell}_{j-1}$, and $\hat{m}_j$ such that
\begin{equation}\label{eq:zumvjt}
\left(\vect{z}(\hat{\ell}_{j-1}), \vect{u}(\hat{\ell}_{j-1}, \hat{m}_j), \vect{v}(\hat{\ell}_j,\hat{k}_j),\bseq{y}{j}\right) \textnormal{ are jointly typical.}
\end{equation}
If such indices can be found, then it picks the first triple of such indices, outputs $\hat{m}_j$ as its guess for $m_j$, and keeps $\hat{\ell}_{j-1}$ to be used when decoding Block $(j-1)$.

\emph{Analysis.} A number of failure modes must be addressed.
\begin{itemize}
\item Decoding of $\ell_{\lambda-1}$ fails. Since the capacity with
  the aforementioned symbol-by-symbol helper is positive, there exists a
  $\gamma>0$ such that, provided Block $\lambda$ has length at least
  $\gamma n$, the probability of a decoding error in this last block
  can be made to tend to zero as $n\to\infty$. Since we shall
  later choose $\lambda$ very  large, the exact value of $\gamma$
  will not affect the overall rate.
\item When encoding for Block $j$, the encoder and the helper cannot
  find indices satisfying \eqref{eq:vsjt}. Notice that the vectors
  \eqref{eq:vgen} are generated IID according to $P_V$ and
  independently of $\bseq{s}{j}$ and
  $\bseq{z}{j}=\vect{z}(\ell_{j-1})$. It follows that these vectors
  are also independent of $\bseq{t}{j}$, whereas the latter, when
  averaged over the randomly generated codebook, is IID according to
  $P_T$. Hence the probability of this event can be made to vanish as 
  $n\to\infty$ as long as
\begin{equation}\label{eq:RvRtil}
R_v + \tilde{R} > I(V;T).
\end{equation}
\end{itemize}
The following error events all concern the decoding task for Block $j\in\{1,\ldots,\lambda-1\}$. We assume that $\ell_j$ has been correctly decoded in Block $(j+1)$ (so $\hat{\ell}_j=\ell_j$).
\begin{itemize}
\item The chosen indices satisfy \eqref{eq:vsjt}, but
  $\left(\bseq{z}{j}, \bseq{u}{j}, \bseq{v}{j},\bseq{y}{j}\right)$ are
  not jointly typical. By our construction, over the randomly
  generated codebook,
  $\left(\bseq{S}{j},
    \bseq{Z}{j},\bseq{T}{j},\bseq{U}{j},\bseq{X}{j},\bseq{Y}{j}\right)$
  are IID according to $P_{SZTUXY}$; in particular, conditional on 
  any~$\bseq{t}{j}$, the probability of the tuple
  $\left(\bseq{z}{j},\bseq{u}{j},\bseq{y}{j}\right)$ is
  $P_{ZUY|T}^{\times
    n}\left(\bseq{z}{j},\bseq{u}{j},\bseq{y}{j}\middle |
    \bseq{t}{j}\right)$. It then follows by the Markov Lemma
  \cite[Lemma 12.1]{elgamalkim11} that, given
  jointly typical $\left(\bseq{v}{j},\bseq{t}{j}\right)$, the
  probability that $\left(\bseq{Z}{j},\bseq{U}{j},\bseq{Y}{j}\right)$
  are jointly typical with $\bseq{v}{j}$ tends to one as $n\to\infty$.

\item There exists some $k' \neq k_j$ such that $\left( \vect{v}
    (\ell_j,k'),\bseq{y}{j}\right)$ are jointly typical.\footnote{This
    will cause an error not in decoding $m_j$, but in decoding
    $\ell_{j-1}$, which will (very likely) cause decoding errors in
    Blocks $j-1$ to $1$.} By our construction, $\vect{v}(\ell_j,k')$
  is generated IID according to $P_V$ and independently of
  $\bseq{y}{j}$. It then follows that the probability of this error
  event can be made to approach zero as $n\to\infty$ as long as
\begin{equation}\label{eq:Rtil}
\tilde{R} < I(V;Y).
\end{equation}
Note that, in the joint distribution \eqref{eq:BMjoint}, $V\markov T \markov Y$ forms a Markov chain. Therefore \eqref{eq:RvRtil} and \eqref{eq:Rtil} together imply that
\begin{equation}\label{eq:Rv}
R_v > I(V;T|Y).
\end{equation}
Conversely, given $R_v$ satisfying \eqref{eq:Rv}, there exists a choice of $\tilde{R}$ to satisfy both \eqref{eq:RvRtil} and \eqref{eq:Rtil}.

\item There exists some $m'\neq m_j$ such that $\left(\vect{z}(\ell_{j-1}), \vect{u}(\ell_{j-1}, m'), \vect{v}(\ell_j,k_j),\bseq{y}{j}\right)$ are jointly typical. As discussed above, with high probability $\left(\vect{z}(\ell_{j-1}), \vect{v}(\ell_j,k_j),\bseq{y}{j}\right)$ are jointly typical. By our construction, $\vect{u}(\ell_{j-1}, m')$ is generated with probability $P_{U|Z}^{\times n} \left(\vect{u}(\ell_{j-1}, m')\middle| \vect{z}(\ell_{j-1})\right)$, hence the probability of a so-generated sequence being jointly typical with  $\left(\vect{z}(\ell_{j-1}), \vect{v}(\ell_j,k_j),\bseq{y}{j}\right)$ is approximately $2^{-n I(U;V,Y|Z)}$. It follows that the probability of this error can be made to vanish as $n\to\infty$ provided
\begin{equation}
R < I(U;V,Y|Z) = I(U;Y|V,Z),
\end{equation}
where the equality follows because, in the joint distribution \eqref{eq:BMjoint}, $U\markov Z \markov V$ forms a Markov chain.

\item There exist some $\ell'\neq \ell_{j-1}$ and $m'\neq m_j$ such that $\left( \vect{z}(\ell'),\vect{u}(\ell',m'), \vect{v}(\ell_j,k_j),\bseq{y}{j}\right)$ are jointly typical. With high probability,  $\left(\vect{v}(\ell_j,k_j),\bseq{y}{j}\right)$ are jointly typical, and, independently, $\left( \vect{z}(\ell'),\vect{u}(\ell',m')\right)$ are drawn IID according to $P_{ZU}$. Therefore the probability that these vectors are jointly typical for any pair $(\ell', m')$ is approximately $2^{-n I(U,Z;V,Y)}$. Consequently, the probability of this type of error can be made to vanish as $n\to\infty$ provided
\begin{equation}
R+R_v < I(U,Z;V,Y).
\end{equation}
\end{itemize}

Summarizing the above analyses, we conclude that, for the block-Markov
scheme to succeed with high probability, it suffices that the rate $R$ be smaller than
\begin{equation} \label{eq:amos_bm_is_good}
  \min \bigl\{ I(U;Y|V,Z),\,\, I(U,Z;V,Y)-I(V;T|Y)\bigr\}.
\end{equation}

We now return to our assertion regarding the existence of a
symbol-by-symbol helper of positive capacity and show that
\eqref{eq:amos_bm_is_good} does not exceed the capacity with a
message-cognizant helper~\eqref{eq:hkm}, let alone the capacity when
perfect state information is available to both encoder and
decoder. Recall that this will allow us to dispense with the
assumption that there exists an effective symbol-by-symbol helper,
which we need in Block~$\lambda$.

To show the desired inequality, consider the second term in the
minimizaiton in \eqref{eq:amos_bm_is_good}:
\begin{IEEEeqnarray}{rCl}
\IEEEeqnarraymulticol{3}{l}{
I(U,Z;V,Y)-I(V;T|Y)}\nonumber\\*
 & = & I(U,Z;Y) + I(U,Z;V|Y) - H(V|Y) + H(V|T) \label{eq:last1} \IEEEeqnarraynumspace \\
& = & I(U,Z;Y) - H(V|U,Z,Y) + H(V|T)\\
& \le & I(U,Z;Y),\label{eq:last2}
\end{IEEEeqnarray}
where \eqref{eq:last1} holds because $V\markov T \markov Y$ forms a Markov chain; and \eqref{eq:last2} because $V\markov T \markov (U,Z,Y)$ forms a Markov chain. If we define 
\begin{equation}
U' \triangleq (U,Z),
\end{equation}
then the joint distribution of $(U',X,Y,S,T)$ has the form \eqref{eq:hkmjoint} with $U'$ replacing $U$, therefore the right-hand side of \eqref{eq:last2} cannot exceed \eqref{eq:hkm}.

We have now completed the proof of the following result:

\medskip

\begin{theorem}\label{thm:blockmarkov}
Given any joint distribution of the form~\eqref{eq:BMjoint}, the
described block-Markov coding scheme can achieve any rate up to \eqref{eq:amos_bm_is_good}.
\end{theorem}

\medskip

\begin{claim}
  Maximizing~\eqref{eq:amos_bm_is_good} over all joint distributions
  of the form~\eqref{eq:BMjoint} yields an achievable rate that is at
  least as high as the capacity with the best symbol-by-symbol
  helper. In some cases it is strictly higher.
  \end{claim}
  
\medskip

  \begin{IEEEproof}
    Choosing $Z$ and $V$ null (deterministic)
    reduces~\eqref{eq:amos_bm_is_good} to $I(U;Y)$, so the
    optimization over $P_{U}$ leads to the capacity with the
    symbol-by-symbol helper $P_{T|S}$.  

    As for an example where the maximum of~\eqref{eq:amos_bm_is_good}
    is higher than the capacity of the best symbol-by-symbol helper,
    consider Example~\ref{ex:from_hell} with the following choices
    (where equalities are with probability one):
\begin{subequations}\label{eq:bungi100}
\begin{IEEEeqnarray}{rCl}
Z & \sim & \textnormal{uniform on }\{0,1\}\\
T & = & S^{(Z)}\\
U & = & (Z,C)\textnormal{ with }C\textnormal{ uniform on }\{0,1\}^\eta\\
X & = & (Z,T,C)\\
V & = & T = S^{(Z)}.
\end{IEEEeqnarray}
\end{subequations}
The output is then
\begin{equation}
Y = \begin{cases} (Z, C, \tilde{C}), & S^{(Z)}=0\\ (Z, \tilde{C},C), & S^{(Z)}=1, \end{cases}
\end{equation}
where $\tilde{C}$ is independent of all other random variables. We can then compute
\begin{IEEEeqnarray}{rCl}
I(U;Y|V,Z) & = & \eta\\
I(U,Z;V,Y) & = & \eta+1\\
I(V;T|Y) & = & 1.
\end{IEEEeqnarray}
The choice~\eqref{eq:bungi100} thus results
in~\eqref{eq:amos_bm_is_good} being $\eta$, which, by
Claim~\ref{cl:from_hell}, is higher than the capacity of any
symbol-by-symbol helper.
\end{IEEEproof}

\section{Concluding Remarks}\label{sec:conclusion}

Revealing the state of an SD-DMC to the
encoder strictly causally does not increase capacity. The intuitive
explanation that is usually given for this is that, because the channel
and state are memoryless, the past states tell the encoder nothing
about the channel's present behavior, and---while possibly useful to
the decoder---it is more efficient for the encoder to convey fresh
information than past states.
So why, as Example~\ref{ex:from_hell} shows, are symbol-by-symbol
helpers suboptimal?

A clue to this might be offered by the coding scheme we proposed for
this example and by the general block-Markov scheme that builds on it.
The idea behind both is that at time~$i$ the encoder conveys to the
decoder some information about the past states that it has learned via
past assistance and \emph{that is known to the helper} (who knows
$S^{i-1}$ and consequently all the past assistance).
Since the helper knows this information about $S^{i-1}$, this
information plays a role similar to that of a message that is known to
the helper. Since revealing the message to the helper can increase
capacity (Example~\ref{ex:hkm}), it can be more efficient for
the encoder to send this information about $S^{i-1}$
than to send fresh information, which the helper does not know.

Such a scheme---where the encoder tells the decoder about past
states---makes no sense if the decoder has full state
information. This is congruent with the optimality of
symbol-by-symbol helpers when the receiver has perfect state
information
(Theorem~\ref{thm:psi_at_decoder}).

Such a scheme also makes no sense when the helper is cognizant of the
message. In this case conveying fresh information is preferable to
conveying information about the past states, because, in comparing the
two approaches, both are done with the helper's knowledge of what is
being conveyed, so the playing field is level. This provides intuition
for Theorem~\ref{thm:hkm}, which states that, when the helper is
cognizant of the message, symbol-by-symbol helpers are optimal.


Finally, we note that, although the assistance provided in
Example~\ref{ex:from_hell} and in the block-Markov scheme depends on
past states, it does not provide the encoder with any new information
about the past states; it only provides the encoder information about
the current state. The past states, however, determine \emph{which
  information about the current state} is provided to the
encoder. Indeed, in this example, even if the past states were
provided to the encoder perfectly, symbol-by-symbol helpers would
still be suboptimal. (By Theorem~\ref{thm:sbs}, the capacity with a
symbol-by-symbol helper is unchanged when past states are provided to
the encoder, so revealing the past states perfectly to the encoder
would not allow the symbol-by-symbol helpers to catch up with the
block-Markov scheme.)

\appendix
In this appendix we complete the proof of Claim~\ref{cl:from_hell} by
analyzing the helper $T=S^{(0)}$. To this end, we view $S^{(0)}$ as a
meta state, which is known causally to the encoder but not to the
decoder. The capacity is then the maximum of $I(U;Y)$, where $U$ takes
values in the set of Shannon strategies that map $S^{(0)}$ to $X$
\cite{shannon58}. Denote a generic Shannon strategy $u$ as
\begin{equation}
u = (a^{(0)},b^{(0)},c^{(0)},a^{(1)},b^{(1)},c^{(1)}),
\end{equation}
indicating that $u$ maps $S^{(0)}=0$ to the channel input
$(a^{(0)},b^{(0)},c^{(0)})$ and $S^{(0)}=1$ to
$(a^{(1)},b^{(1)},c^{(1)})$. The conditional law of $Y$ given $U$ is then
\begin{IEEEeqnarray}{l}
P_{Y|U}(y|u)  =  \sum_{s^{(1)} \in \{0,1\}} \frac{1}{2} \, W\bigl(y \big| x = (a_{0},b_{0},c_{0}), s = (0,s^{(1)}) \bigr) \nonumber \\*
\quad\,{} + \> \sum_{s^{(1)} \in \{0,1\}} \frac{1}{2} \, W\bigl(y \big| x = (a_{1},b_{1},c_{1}), s = (1,s^{(1)}) \bigr).  \IEEEeqnarraynumspace
\end{IEEEeqnarray}

The duality-based upper bound
\cite[Thm.~5.1]{lapidothmoser03_3} states that any choice of a distribution
$Q$ on~$\set{Y}$ leads to an upper bound on the capacity
\begin{equation}
C \le \max_u D\big(P_{Y|U}(\cdot|u) \big\| Q\big).
\end{equation}
Our choice of $Q$ is one under which $A'$, $D^{(0)}$, and $D^{(1)}$ are independent, with $A'$ being Bernoulli ($\delta$) (with $\delta < 1/2$ specified later) and with both $D^{(0)}$ and $D^{(1)}$ uniform:
\begin{equation}
Q(a',d^{(0)},d^{(1)}) = \begin{cases} (1-\delta) \, 2^{-2\eta},& a'=0\\ \delta\, 2^{-2\eta},& a'=1. \end{cases}
\end{equation}

We next analyze $D\big(P_{Y|U}(\cdot|u) \big\| Q\big)$ for different strategies~$u$.
\begin{itemize}
\item Consider any $u$ with $a^{(0)}=a^{(1)}=0$, $b^{(0)}=0$, and
$b^{(1)}=1$. Such a $u$ guarantees that $B = S^{(A)} = S^{(0)}$, and
\begin{equation}
Y = \begin{cases} (0, c^{(0)}, \tilde{D} ) & \textnormal{ w.p. }1/2\\ (0, \tilde{D}, c^{(1)}) & \textnormal{ w.p. } 1/2, \end{cases}
\end{equation}
where $\tilde{D}$ is uniform over $\{0,1\}^\eta$ and independent of
everything else. Hence $P_{Y|U}(y|u)$ equals $2^{-\eta}$ for $y=
(0,c^{(0)},c^{(1)})$, and equals $2^{-\eta-1}$ for the other
$(2^{\eta+1}-2)$ outcomes of positive probability. (Outputs of the
form $(0,\kappa,\kappa')$ where $\kappa \neq c^{(0)}$ and $\kappa' \neq
c^{(1)}$ have zero probability.) It then follows that
\begin{IEEEeqnarray}{rCl}
\IEEEeqnarraymulticol{3}{l}{
D\big(P_{Y|U}(\cdot|u) \big\| Q\big)}\nonumber\\*
\quad & = & 2^{-\eta} \log\frac{2^{-\eta}}{(1-\delta) 2^{-2\eta}} + (1-2^{-\eta}) \log\frac{2^{-\eta-1}}{(1-\delta)2^{-2\eta}}\nonumber\\*\\
& = & \eta + 2^{-\eta} - 1 + \log\frac{1}{1-\delta}.
\end{IEEEeqnarray}

\item Now consider $u$ with $a^{(0)}=a^{(1)}=0$, $b^{(0)}=1$, and $b^{(1)}=0$. For this $u$, we always have $B\neq S^{(A)}$, so $Y$ conditional on $u$ is always uniform, and
\begin{IEEEeqnarray}{rCl}
D\big(P_{Y|U}(\cdot|u) \big\| Q\big) & = & D\big( P_{A'|U}(\cdot|u) \big\| Q_{A'} \big) \\
& = & D\big(\textnormal{Bern}(1/2) \big\| \textnormal{Bern}(\delta)\big) \IEEEeqnarraynumspace \\
& \le & \log \frac{1}{\delta}.
\end{IEEEeqnarray}

\item Consider $u$ with $a^{(0)} = a^{(1)} = 0$ and
  $b^{(0)} = b^{(1)} = 0$ (so the meta state is nearly
  ignored). Conditional on such a $u$, $B=S^{(A)}$ when $S^{(0)}=0$
  and $B \neq S^{(A)}$ when $S^{(0)}=1$, so
\begin{equation}
Y =\begin{cases} (0,c^{(0)}, \tilde{D}^{(1)}) & \textnormal{ w.p. }1/2\\
(\tilde{A},\tilde{D}^{(0)},\tilde{D}^{(1)}) & \textnormal{ w.p. } 1/2,\end{cases}
\end{equation}
where $\tilde{A},\tilde{D}^{(0)},\tilde{D}^{(1)}$ are all uniform and
independent of everything else.  In the first case above, we have a
probability of $2^{-\eta}$ on each of the $2^\eta$ realizations of $Y$
of positive probability; and in the second case, we have a probability
of $2^{-2\eta-1}$ on each outcome of positive probability. By
convexity of relative entropy, we can upper-bound
$D\big(P_{Y|U}(\cdot|u) \big\| Q\big)$ by the weighted sum of the
relative entropies resulting from these two cases (in the second case
it is $D\big(\textnormal{Bern}(1/2) \big\|
\textnormal{Bern}(\delta)\big)$):
\begin{IEEEeqnarray}{rCl}
\IEEEeqnarraymulticol{3}{l}{
D\big(P_{Y|U}(\cdot|u) \big\| Q\big)
}\nonumber\\*
 & \le & \frac{1}{2} \log \frac{2^{-\eta}}{(1-\delta)2^{-2\eta}}+ \frac{1}{2}D\big(\textnormal{Bern}(1/2) \big\| \textnormal{Bern}(\delta)\big)\IEEEeqnarraynumspace \\
& \le & \frac{\eta}{2}+\log \frac{1}{\delta}.\label{eq:app175}
\end{IEEEeqnarray}

One can easily verify that the same bound holds for $a^{(0)} = a^{(1)} = 0$ and $b^{(0)} = b^{(1)} = 1$.

\item Consider $u$ with $a^{(0)} = a^{(1)} = 1$ and
  $b^{(0)}=b^{(1)}=0$ (so the meta state is again nearly
  ignored). Conditional such a $u$, there is a probability of $1/2$
  that $B \neq S^{(A)} = S^{(1)}$, and
\begin{equation}
Y = \begin{cases} (\tilde{A},\tilde{D}^{(0)},\tilde{D}^{(1)}) & \textnormal{ w.p. } 1/2\\
(1,c^{(0)},\tilde{D}^{(1)}) &  \textnormal{ w.p. } 1/4\\
(1,c^{(1)},\tilde{D}^{(1)}) &  \textnormal{ w.p. } 1/4, \end{cases}
\end{equation}
where $\tilde{A},\tilde{D}^{(0)},\tilde{D}^{(1)}$ are all uniform and independent of everything else. Again using convexity of relative entropy we have
\begin{IEEEeqnarray}{rCl}
\IEEEeqnarraymulticol{3}{l}{
D\big(P_{Y|U}(\cdot|u) \big\| Q\big)
}\nonumber\\*  \qquad
 & \le & \frac{1}{2} D\big(\textnormal{Bern}(1/2) \big\| \textnormal{Bern}(\delta)\big) \nonumber\\*
  & & {} + \frac{1}{4} \log \frac{2^{-\eta}}{\delta 2^{-2\eta}} + \frac{1}{4} \log \frac{2^{-\eta}}{\delta 2^{-2\eta}}\IEEEeqnarraynumspace \\
& \le & \frac{\eta}{2} + \log \frac{1}{\delta}. \label{eq:app178}
\end{IEEEeqnarray}
By similar analysis, the bound \eqref{eq:app178} can be shown to hold for all $u$ with $a^{(0)}=a^{(1)}=1$. 

\item Next consider $u$ with $a^{(0)}=0$, $a^{(1)}=1$, $b^{(0)}=b^{(1)}=0$. When $S^{(0)}=0$ we always have $B=S^{(A)}=S^{(0)}$; but, when $S^{(0)}=1$, $B$ is independent of $S^{(A)}=S^{(1)}$. The output is thus as follows:
\begin{equation}
Y = \begin{cases} (0,c^{(0)},\tilde{D}^{(1)}) & \textnormal{ w.p. } 1/2\\ (1, c^{(0)},\tilde{D}^{(1)}) & \textnormal{ w.p. } 1/4\\(\tilde{A},\tilde{D}^{(0)},\tilde{D}^{(1)}) & \textnormal{ w.p. } 1/4,\end{cases}
\end{equation}
where $\tilde{A},\tilde{D}^{(0)},\tilde{D}^{(1)}$ are all uniform and independent of everything else. Using convexity of relative entropy we have
\begin{IEEEeqnarray}{rCl}
\IEEEeqnarraymulticol{3}{l}{
D\big(P_{Y|U}(\cdot|u) \big\| Q\big)
}\nonumber\\* \qquad
 & \le & \frac{1}{2} \log \frac{2^{-\eta}}{(1-\delta) 2^{-2\eta}} + \frac{1}{4}  \log \frac{2^{-\eta}}{\delta 2^{-2\eta}} \nonumber\\*
 & & {} + \frac{1}{4} D\big(\textnormal{Bern}(1/2) \big\| \textnormal{Bern}(\delta)\big)   \\
& \le & \frac{3\eta}{4} + \log\frac{1}{\delta}.\label{eq:app182}
\end{IEEEeqnarray}
The same bound holds for $a^{(0)}=0$, $a^{(1)}=1$, $b^{(0)}=0$, $b^{(1)}=1$, as well as for $a^{(0)}=1$, $a^{(1)}=0$, $b^{(1)}=1$ (and $b^{(0)}=0$ or $1$).

\item Finally, consider $a^{(0)}=0$, $a^{(1)}=1$, $b^{(0)}=1$, and $b^{(1)}=0$. (The bound will also apply to $a^{(0)}=0$, $a^{(1)}=1$, $b^{(0)}=b^{(1)}=1$, and to $a^{(0)}=1$, $a^{(1)}=0$, $b^{(1)}=0$, $b^{(0)}=0$ or $1$.) In this case, when $S^{(0)}=0$, we always have $B\neq S^{(A)}=S^{(0)}$; and when $S^{(0)}=1$, we have that $B$ is independent of $S^{(A)}=S^{(1)}$. Consequently,
\begin{equation}
Y = \begin{cases}(1, c^{(1)},\tilde{D}^{(1)}) & \textnormal{ w.p. } 1/4\\(\tilde{A},\tilde{D}^{(0)},\tilde{D}^{(1)}) & \textnormal{ w.p. } 3/4,\end{cases}
\end{equation}
where $\tilde{A},\tilde{D}^{(0)},\tilde{D}^{(1)}$ are all uniform and independent of everything else. We then have
\begin{IEEEeqnarray}{rCl}
\IEEEeqnarraymulticol{3}{l}{
D\big(P_{Y|U}(\cdot|u) \big\| Q\big)
}\nonumber\\*
\quad & \le & \frac{1}{4}  \log \frac{2^{-\eta}}{\delta 2^{-2\eta}} + \frac{3}{4} D\big(\textnormal{Bern}(1/2) \big\| \textnormal{Bern}(\delta)\big) \IEEEeqnarraynumspace \\
& \le & \frac{\eta}{4} + \log\frac{1}{\delta}.
\end{IEEEeqnarray}
\end{itemize}

Summarizing all above cases, the capacity with the helper $T=S^{(0)}$ is upper-bounded as
\begin{equation}
C \le \max\left\{ \eta + 2^{-\eta} - 1 + \log\frac{1}{1-\delta}, \,\,\frac{3\eta}{4} + \log\frac{1}{\delta}\right\}. \label{eq:duality_final}
\end{equation}
If we choose $\delta = 1/4$, then the bound becomes
\begin{equation}
C \le \max \left\{ \eta+2^{-\eta} -\log\frac{3}{2},\,\, \frac{3\eta}{4} + 2\right\},
\end{equation}
which is less than $\eta$ if $\eta>8$. This completes the proof.

When $\eta$ is very large, we can choose $\delta$ to be close to zero, and the best bound obtained from \eqref{eq:duality_final} will be approximately $\eta-1$. In fact, $\eta-1$ is indeed approximately the best performance with a symbol-by-symbol helper when $\eta$ is very large.

\end{document}